\documentclass[12pt]{article}
\usepackage{graphicx,psfrag}

\newcounter{multieqs}

\newcommand{\be}{\begin{equation}}
\newcommand{\ee}{\end{equation}}
\newcommand{\eq}[1]{(\ref{#1})}

\newcommand{\rf}[1]{(\ref{#1})}

\def\bd{\begin{document}}
\def\ed{\end{document}}
\def\nn{\nonumber}
\def\bea{\begin{eqnarray}}
\def\eea{\end{eqnarray}}
\let\bm=\bibitem
\let\la=\label

\def\npb#1#2#3{Nucl. Phys. {\bf{B#1}} (#2) #3}
\def\plb#1#2#3{Phys. Lett. {\bf{B#1}} (#2) #3}
\def\prl#1#2#3{Phys. Rev. Lett. {\bf{#1}} (#2) #3}
\def\prd#1#2#3{Phys. Rev. {D \bf{#1}} (#2) #3}
\def\cmp#1#2#3{Comm. Math. Phys. {\bf{#1}} (#2) #3}
\def\cqg#1#2#3{Class. Quantum Grav. {\bf{#1}} (#2) #3}
\def\nppsa#1#2#3{Nucl. Phys. B (Proc. Suppl.) {\bf{#1A}} (#2) #3}
\def\ap#1#2#3{Ann. of Phys. {\bf{#1}} (#2) #3}
\def\ijmp#1#2#3{Int. J. Mod. Phys. {\bf{A#1}} (#2) #3}
\def\rmp#1#2#3{Rev. Mod. Phys. {\bf{#1}} (#2) #3}
\def\mpla#1#2#3{Mod. Phys. Lett. {\bf A#1} (#2) #3}
\def\jhep#1#2#3{JHEP {\bf #1} (#2) #3}
\def\atmp#1#2#3{Adv. Theor. Math. Phys. {\bf #1} (#2) #3}

\newcommand{\EQ}[1]{\begin{equation} #1 \end{equation}}
\newcommand{\AL}[1]{\begin{subequations}\begin{align} #1 \end{align} 
\end{subequations}}
\newcommand{\SP}[1]{\begin{equation}\begin{split} #1 \end{split}\end{equation}}
\newcommand{\ALAT}[2]{\begin{subequations}\begin{alignat}{#1} #2 
\end{alignat}\end{subequations}}

\def\beqa{\begin{eqnarray}} 
\def\eeqa{\end{eqnarray}} 

\def\beq{\begin{equation}} 
\def\eeq{\end{equation}}

\def\sst{\scriptscriptstyle}
\def\thetabar{\bar\theta}
\def\Tr{{\rm Tr}}
\def\one{\mbox{1 \kern-.59em {\rm l}}}

\def\a{\alpha}      \def\da{{\dot\alpha}}  
\def\b{\beta}       \def\db{{\dot\beta}}  
\def\c{\gamma}  \def\C{\Gamma}  \def\dc{{\dot\gamma}}  
\def\d{\delta}  \def\D{\Delta}  \def\ddt{\dot\delta}  
\def\e{\epsilon}        \def\vare{\varepsilon}  
\def\f{\phi}    \def\F{\Phi}    \def\vvf{\f}  
\def\h{\eta}  
\def\k{\kappa}  
\def\l{\lambda} \def\L{\Lambda}  
\def\m{\mu} \def\n{\nu}  
\def\o{\omega}  
\def\p{\pi} \def\P{\Pi}  
\def\r{\rho}  
\def\s{\sigma}  \def\S{\Sigma}  
\def\t{\tau}  
\def\th{\theta} \def\Th{\Theta} \def\vth{\vartheta}  
\def\X{\Xeta}  
\def\z{\zeta}  
  
\def\cA{{\cal A}} \def\cB{{\cal B}} \def\cC{{\cal C}}  
\def\cD{{\cal D}} \def\cE{{\cal E}} \def\cF{{\cal F}}  
\def\cG{{\cal G}} \def\cH{{\cal H}} \def\cI{{\cal I}}  
\def\cJ{{\cal J}} \def\cK{{\cal K}} \def\cL{{\cal L}}  
\def\cM{{\cal M}} \def\cN{{\cal N}} \def\cO{{\cal O}}  
\def\cP{{\cal P}} \def\cQ{{\cal Q}} \def\cR{{\cal R}}  
\def\cS{{\cal S}} \def\cT{{\cal T}} \def\cU{{\cal U}}  
\def\cV{{\cal V}} \def\cW{{\cal W}} \def\cX{{\cal X}}  
\def\cY{{\cal Y}} \def\cZ{{\cal Z}}

\def\ua{\underline{\alpha}}  
\def\ub{\underline{\phantom{\alpha}}\!\!\!\beta}  
\def\uc{\underline{\phantom{\alpha}}\!\!\!\gamma}  
\def\um{\underline{\mu}}  
\def\ud{\underline\delta}  
\def\ue{\underline\epsilon}  
\def\una{\underline a}\def\unA{\underline A}  
\def\unb{\underline b}\def\unB{\underline B}  
\def\unc{\underline c}\def\unC{\underline C}  
\def\und{\underline d}\def\unD{\underline D}  
\def\une{\underline e}\def\unE{\underline E}  
\def\unf{\underline{\phantom{e}}\!\!\!\! f}\def\unF{\underline F}  
\def\unm{\underline m}\def\unM{\underline M}  
\def\unn{\underline n}\def\unN{\underline N}  
\def\unp{\underline{\phantom{a}}\!\!\! p}\def\unP{\underline P}  
\def\unq{\underline{\phantom{a}}\!\!\! q}  
\def\unQ{\underline{\phantom{A}}\!\!\!\! Q}  
\def\unH{\underline{H}}  
  
\def\As {{A \hspace{-6.4pt} \slash}\;}  
\def\bs {{b \hspace{-6.4pt} \slash}\;}  
\def\Ds {{D \hspace{-6.4pt} \slash}\;}  
\def\ds {{\del \hspace{-6.4pt} \slash}\;}  
\def\ss {{\s \hspace{-6.4pt} \slash}\;}  
\def\ks {{ k \hspace{-6.4pt} \slash}\;}  
\def\ps {{p \hspace{-6.4pt} \slash}\;}  
\def\pas {{{p_1} \hspace{-6.4pt} \slash}\;}  
\def\pbs {{{p_2} \hspace{-6.4pt} \slash}\;}  
  
\def\Rh{\hat{R}}  
\def\Gh{\hat{G}}
\def\Fh{\hat{F}}
\def\Hh{\hat{H}}
\def\Ah{\hat{A}}
\def\Bh{\hat{B}}
\def\sh{\hat{s}}  
\def\xh{\hat{x}}  
\def\yh{\hat{y}}  
\def\gh{\hat{g}}  
\def\xih{\hat{\xi}}  
  
\def\psit{\tilde{\psi}}  
\def\Psit{\tilde{\Psi}}   
\def\Psibt{\tilde{\bar{Psi}}}  

\def\Phit{\tilde{\Phi}}   
\def\Phitb{\overline{\tilde{Phi}}}  
\def\tht{\tilde{\th}}  
\def\xit{\tilde{\xi}}
   
\def\At{\tilde{A}}  
\def\Qt{\tilde{Q}}  
\def\Rt{\tilde{R}}  
\def\Nt{\tilde{N}}  
\def\Gt{\tilde{G}}
 
\def\at{\tilde{a}}  
\def\st{\tilde{s}}  
\def\ft{\tilde{f}}  
\def\pt{\tilde{p}}  
\def\qt{\tilde{q}}  
\def\vt{\tilde{v}}  
\def\nt{\tilde{n}}  
  
\def\delb{\overline{\partial}}  
\def\thb{\overline{\theta}}
\def\mub{{\overline \mu}}

\def\lamb{{\overline \l}}
\def\psib{{\overline \psi}}
\def\sb{{\overline \sigma}}
\def\xib{{\overline \xi}}
\def\chib{{\overline \chi}}

\def\Phib{\overline{\Phi}}
\def\Lamb{\overline{\Lambda}}

\def\Ab{{\overline A}} \def\Bb{{\overline B}} \def\Cb{{\overline C}}  
\def\Db{{\overline D}} \def\Eb{{\overline E}} \def\Fb{{\overline F}}  
\def\Gb{{\overline G}} \def\Hb{{\overline H}} \def\Ib{{\overline I}}  
\def\Jb{{\overline J}} \def\Kb{{\overline K}} \def\Lb{{\overline L}}  
\def\Mb{{\overline M}} \def\Nb{{\overline N}} \def\Ob{{\overline O}}  
\def\Pb{{\overline P}} \def\Qb{{\overline Q}} \def\Rb{{\overline R}}  
\def\Sb{{\overline S}} \def\Tb{{\overline T}} \def\Ub{{\overline U}}  
\def\Vb{{\overline V}} \def\Wb{{\overline W}} \def\Xb{{\overline X}}  
\def\Yb{{\overline Y}} \def\Zb{{\overline Z}}  

\def\fb{{\overline f}}
\def\gb{{\overline g}}
\def\mb{{\overline m}}
\def\lb{{\overline l}}
\def\yb{{\overline y}}

\def\ba{{\bf a}} 
\def\bk{{\bf k}}  
\def\bl{{\bf l}}  
\def\bp{{\bf p}}  
\def\bq{{\bf q}}  
\def\br{{\bf r}}
\def\bt{{\bf t}}
\def\bu{{\bf u}}
\def\bv{{\bf v}}
\def\bx{{\bf x}}  
\def\by{{\bf y}}  
\def\bR{{\bf R}}  
\def\bV{{\bf V}}

\def\bone{{\bf 1}}  

\def\va{{\vec a}}
\def\vp{{\vec p}}
\def\vq{{\vec q}}
\def\vx{{\vec x}}
\def\vu{{\vec u}}
\def\vv{{\vec v}}

\def\vs{{\vec \sigma}}
\def\vtau{{\vec \tau}}

\newcommand{\ov}[1]{\overrightarrow{#1}}
  
\def\ddt{\dot\delta}  
  
\def\pa{\partial} \def\del{\partial}  
\def\xx{\times}  
\def\uno{\mbox{1 \kern-.59em {\rm l}}}    
  
\def\trp{^{\top}}  
\def\inv{^{-1}}  
\def\dag{{^{\dagger}}}  
\def\pr{^{\prime}}  
  
\def\rar{\rightarrow}  
\def\lar{\leftarrow}  
\def\lrar{\leftrightarrow}  
  
\newcommand{\0}{\,\!}      
\def\one{1\!\!1\,\,}  
\def\im{\imath}  
\def\jm{\jmath}  
  
\newcommand{\tr}{\mbox{tr}}  
\newcommand{\slsh}[1]{/ \!\!\!\! #1}  
  
\def\vac{|0\rangle}  
\def\lvac{\langle 0|}  
  
\def\hlf{\frac{1}{2}}  
\def\ove#1{\frac{1}{#1}}  

\def\Box{\square}  
\def\ZZ{\mathbb{Z}}  
\def\bb#1{{\bf #1}}  
\def\bcomment#1{}  
\def\bfhat#1{{\bf \hat{#1}}}  
\def\VEV#1{\left\langle #1\right\rangle}  

\newcommand{\ex}[1]{{\rm e}^{#1}} \def\ii{{\rm i}}  

\newcommand{\lrbrk}[1]{\left(#1\right)}
\newcommand{\sfrac}[2]{{\textstyle\frac{#1}{#2}}}
 
\def\stw{{\sqrt{2}}}

\def\rf {{\rm f}}
\def\ri {{\rm i}}
\def\rs {{\scriptscriptstyle \rm S}}
\def\rt {{\scriptscriptstyle \rm T}}

\def\rQ {{\scriptscriptstyle \rm \cQ}}
\def\rR {{\scriptscriptstyle \rm \cR}}

\def\cQb{{\cal \Qb}}
\def\cRb{{\cal \Rb}}
\def\cWb{{\cal \Wb}}

\def\fd {{\rm N}}
\def\afd {{\overline{\rm N}}}

\def\Eu {{\mbox{\tiny\it E}}}

\newcommand{\sub}[1]{\mbox{\tiny \it #1}}

\def\csch{{\rm csch}}
\def\oneone{\rlap 1\mkern4mu{\rm l}}
\def\im{{{\rm i}}}

\def\1{{\sst{(1)}}}
\def\2{{\sst{(2)}}}
\def\3{{\sst{(3)}}}

\def\Fth{\hat{\tilde{F}}}

\def\xih{\hat{\xi}}  
\def\chih{\hat{\chi}}
\def\psih{\hat{\psi}}
\def\eh{\hat{\e}} 
\def\g{\gamma}  \def\G{\Gamma}  \def\dg{{\dot\gamma}}

\font\myBB=msbm10 at 18pt
\def\BB#1{\hbox{\myBB#1}}

\begin{document}
\hfill{{\tt hep-th/0611325}}

\vspace{20pt}

\begin{center}

{\Large \bf  On Black Ring  with a Positive Cosmological Constant }

\vspace{30pt}

{\bf Chong-Sun Chu, Shou-Huang Dai }

\vspace{15pt}
{\small \em
\begin{itemize}
\item[]
Centre for Particle Theory and Department of Mathematics,
University of Durham, \\Durham, DH1 3LE, UK.
\end{itemize}
}

\vskip .1in {\small \sffamily chong-sun.chu@durham.ac.uk, 
shou-huang.dai@durham.ac.uk }


\vspace{50pt}
{\bf Abstract}

\end{center}

We consider black ring with a cosmological constant 
in the five dimensional $\cN=4$ de Sitter supergravity theory. 
Our solution preserves
half of the de Sitter supersymmetries and has one rotation symmetry. 
Unlike the flat case, there is no angular momentum and the stability against
gravitational self-attraction is balanced by the cosmological repulsion due to
the cosmological constant.
Our solution describes a singular black ring since although
it has horizons of topology
$S^1\times S^2$, the horizons are singular.  Despite the singularity, our
solution displays some interesting regular physical properties:
it  carries a dipole charge and
this charge contributes to the first law of thermodynamics;
it has an entropy and mass  which conform to the entropic
N-bound proposal and the maximal mass conjecture. 
We conjecture that the
Gregory-Laflamme instability leads to  a resolution of the singularity
and results in a regular black ring.
\setcounter{page}0
\newpage

\section{Introduction}
 
A remarkable black hole solution, the black ring, admitting a horizon of
non-spherical topology was discovered by Emparan and
Reall \cite{br} in 2001. This solution satisfies the vacuum Einstein
equations in five dimensions and has a horizon of $S^1 \times S^2$
topology. The solution is neutral and rotation is needed to prevent 
the ring from gravitational collapse. 
This solution 
has been further generalized to charged black rings \cite{cr}, and
supersymmetric ones  \cite{susy-ring}
using the results of \cite{minimal} which provides a classification 
of all supersymmetric bosonic solutions of minimal supergravity in 
five-dimensions. Black ring exhibits infinite non-uniqueness due to its
unusual dipole charge \cite{dipole-ring}. Unlike the usual black hole hair, 
the dipole charge is not obtained from surface integrals at infinity.
Nevertheless they enter into the first law of thermodynamics
for the black ring \cite{dipole-ring,CH,radu,rogatko}.
Microscopic analysis of entropy of black ring
has been considered in \cite{entropy}.  Relation with integrable
system was noted in \cite{int}.
See \cite{review} for further developments and  a comprehensive review.
 
The black rings constructed so far are asymptotically flat and without a
cosmological constant; 
black ring with a cosmological constant has not been constructed
\footnote{
An interesting  black ring solution 
in $AdS_3 \times S^3$ has been constructed \cite{bena}. In this paper
we are interested in constructing black ring solution in higher
dimensional non-asymptotically flat spacetime.}. 
It is a challenging problem. For applications in 
dS/CFT or AdS/CFT correspondence, it will be very interesting to have
such black rings so that one may investigate how the properties related to
the  nontrivial topology
of horizons are encoded in terms of the dual field theory.
Moreover since our universe is known to have a small
but nonzero positive cosmological constant. 
A 5-dimensional black ring with a
positive cosmological constant may lead to interesting observable
effects in the 4-dimensional low energy world.
The purpose of this paper is to construct a black ring with a positive
cosmological constant and  examine  its properties.
 
The original black ring solution was constructed using the
Kaluza-Klein reduction. A particular solution, the dilatonic C-metric
\cite{dowker} played a central role in the construction. 
The C-metric was first obtained as a special solution 
of the reduced 4-dimensional dilaton-Maxwell-Einstein gravity, and 
substituted into the  Kaluza-Klein  reduction
formula to obtain a 5-dimensional solution. The black ring is then
obtained from a double Wick rotation and by choosing appropriate range
of its coordinates.

The C-metrics describes a pair of uniformly accelerated
black holes in the opposite directions \cite{kw,exact}. It was initially
constructed for the flat spacetime. Generalization to C-metric 
with a cosmological constant 
is not difficult \cite{adsc0} (see also \cite{adsc1,dsc}) and they
are obtained as solutions in the four-dimensional Maxwell-Einstein
gravity with a cosmological constant. To construct a black ring in
five-dimensional spacetime with a cosmological constant,
it is natural to expect that these
C-metrics may play a role. To make use of these C-metrics in a way
similar to the idea of \cite{br}, one need to start with a
five-dimensional theory which under a consistent reduction reduces to
the four-dimensional Maxwell-Einstein gravity with a cosmological
constant (or perhaps with a dilaton also). One may try the conventional
Kaluza-Klein reduction but it is immediately clear that it does not
work. Another possibility is to perform a warp compactification.
Using the braneworld reduction ansatz
\cite{rs2,pope1,pope2},
we will show that when the
cosmological constant is positive, one can  
construct  a supersymmetric solution in the 5-dimensional de Sitter supergravity
\footnote{Our method only works for the 5-dimensional dS case. The
reason will be clear as we explain  our construction in section 3.}
that has horizons of topology  $S^1 \times S^2$. This is our candidate
of a black ring solution.

This solution is singular, however, since  the $S^2$ factor of the horizon
shrinks to zero size as the warp factor vanishes
\footnote{We thank Roberto Emparan and Harvey Reall for pointing out 
this point to us which we missed in our original consideration.}. 
Thus our solution describes a   
singular black ring. Despite the singularity, our 
solution  display very interesting
properties: it has an entropy 
which conform to the the N-bound proposal of Bousso \cite{bousso}.
Moreover it has a negative mass,
suggesting that a maximal mass conjecture
similar to that of \cite{tensor2} may hold in general for spacetime with a
positive cosmological constant. These suggests that there is an underlying 
framework where the singularity of
our solution may get resolved and the solution makes good physical
sense.  Indeed singular horizon is a quite generic feature of
brane world black hole. In  \cite{cgh}, a Schwarzschild black hole is
considered on the brane, it was found that the solution is singular at
the AdS horizon and it was suggested that 
as a result of the Gregory-Laflamme instability \cite{GL}, 
the horizon will pinch off
and a regular ``black cigar'' solution is formed. It is  a
challenging and still open  question to construct this brane world black hole. 
We remark that the outcome of the Gregory-Laflamme instability is still
a matter to be settled. In addition to the original proposal of Gregory
and Laflamme where the horizon pinches off, Horowitz and Maeda \cite{HM} 
have argued that  pinch off cannot happen; and instead, a new horizon is
formed around the "neck region" leading to a new regular horizon.   
In accordance with the scenario of  \cite{HM},
the Gregory-Laflamme instability  for our solution could lead to a 
resolution of the singularity and results in a regular black ring.

The paper is organized as follows. In the next section, we give a brief
review of the braneworld reduction of \cite{pope2}. 
In section 3, we  construct a half supersymmetric solution in the de
Sitter supergravity theory. 
The solution  
has an black ring  horizon as well as a cosmological horizon, 
both of topology $S^1\times S^2$. Despite the singularity of the horizon,  
the surface gravities and the areas (entropies) for these
horizons are finite and well defined. 
The solution has no
global charge, but it has a local dipole charge like the original black ring, 
and this appears in the first law of black hole thermodynamics.  
We discuss the fate of the singular horizon and suggest that the 
Gregory-Laflamme instability to lead to a regular black ring. 
Further discussions are given at the end of section 3 and 
in the section 4.

\section{Braneworld Kaluza-Klein reduction}

Unlike the usual Kaluza-Klein reduction which is based on a
factorizable geometry, the braneworld Kaluza-Klein 
reduction is based on a warp metric. 
Consider an embedding
of the form
\be \label{ansatz-red}
d\sh_{D+1}^2 = dz^2 + f^2 ds_D^2,
\ee
where $f = f(z)$ and the $D$-dimensional metric $ds_D^2$ is Lorentzian.
In this paper, we will use hatted variables to indicate 5-dimensional quantities. 
An important observation is that the
higher dimensional Ricci tensor is simply related to the lower
dimensional one as
\bea
&& \Rh_{zz} = -  \frac{D f''}{f}, \qquad \Rh_{\m z} =0, 
\nn\\
&& \Rh_{\m\n} = \Bigg[ R_{\m\n} +(D-1)(f'' f -f'^2) g_{\m\n} \Bigg] - 
\frac{D f'' }{f} \gh_{\m\n},
\eea
for $\m, \n = 1, \cdots, D$. It follows immediately that for special choices of 
$f = e^{-kz}$, $\cosh(k z)$, $\cos(kz)$, $\sinh(k z)$, one can embed
a lower dimensional constant curvature spacetime within a higher dimensional 
constant curvature spacetime, namely \cite{pope2}
\be\label{red-pope}
\begin{array}{lll}
(i) M_D \subset AdS_{D+1}: & f= e^{-kz}, & \Rh_{z z } = -D k^2, \\
& &  \Rh_{\m\n} = R_{\m\n}  -D k^2 \gh_{\m\n},\\
\\
(ii) AdS_D \subset AdS_{D+1}: & f= \cosh(kz), & \Rh_{z z } = -D k^2, \\
& &  \Rh_{\m\n} = [R_{\m\n} +(D-1)k^2
g_{\m\n}] -D k^2 \gh_{\m\n},\\
\\
(iii) dS_D \subset AdS_{D+1}: & f= \sinh(kz), & \Rh_{z z } = -D k^2 , \\
& & \Rh_{\m\n} = [R_{\m\n} -(D-1)k^2
g_{\m\n}] -D k^2 \gh_{\m\n},\\
\\
(iv) dS_D \subset dS_{D+1}: & f= \cos(kz), & \Rh_{z z } = D k^2,\\
& &  \Rh_{\m\n} = [R_{\m\n} -(D-1)k^2
g_{\m\n}] +D k^2 \gh_{\m\n}.
\end{array}
\ee
An interesting feature of the reduction ansatz \eq{ansatz-red} 
is the change in the cosmological constant.

Moreover it has been demonstrated \cite{pope2} how
these reduction ansatz for the metric can be extended to the other
fields of $\cN=4$ gauge supergravity in five-dimensions to obtain a
consistent reduction of supergravity. The first case is basically the
one considered by Randall and Sundrum. In cases (ii), (iii), one obtains
gauged $\cN=2$, $D=4$ AdS supergravity and, respectively, gauged $\cN=2$, $D=4$
dS supergravity from the gauged $\cN=4$, $D=5$ AdS supergravity. In case
(iv), one obtains gauged $\cN=2$, $D=4$ dS supergravity from the gauged
$\cN=4$, $D=5$ dS supergravity upon reduction.

\subsection{The case of $dS_4 \subset dS_5$}

As we will explain in the next section, 
the case (iv) is the only case which allows one to
construct a black ring solution in a spacetime with a cosmological
constant, therefore we will review  the braneworld reduction for
this case in  more details. The five dimensional theory  to
start with is the five-dimensional gauged $\cN=4$ de Sitter supergravity,
which can be obtained by performing
Kaluza-Klein reduction of type $\mathrm{IIB}^\ast$ theory on $H^5$
\cite{hull0}, where the $\mathrm{IIB}^\ast$ supergravity arises by
performing a T-duality transformation on type IIA on a timelike circle.
As a result the Ramond-Ramond fields have ``wrong-sign'' kinetic terms.
The bosonic fields of the theory are the metric, the dilaton field $\f$,
three $SU(2)$ Yang-Mills potentials $\Ah_\1^i$ ($i=1,2,3$), the $U(1)$
gauge potential $\Bh_\1$ and two 2-form potentials $\Ah_\2^\a$ ($\a
=1,2$) which transform as a charged doublet under the $U(1)$. The
bosonic Lagrangian is
\bea \label{ds5lag}
  \cL_{5(dS)} 
   &=& \Rh \, {\hat *\oneone} - \hlf\,{\hat *d\f}\wedge d\f - 
        \hlf X^4\,\hat * \Gh_\2 \wedge \Gh_\2 
 - \hlf X^{-2}\, (-\hat *\Fh^i_\2 \wedge \Fh^i_\2 + \th_{(\a)} \hat * 
         \Ah^\a_\2 \wedge \Ah^\a_\2) \nn\\
     &&    + \frac{1}{2g} \e_{\a\b}\, \Ah^\a_\2 \wedge d\Ah^\b_\2 
- \hlf \th_{(\a)} \Ah^\a_\2 \wedge \Ah^\a_\2 \wedge \Bh_\1  
       + \hlf \Fh^i_\2 \wedge \Fh^i_\2 \wedge \Bh_\1 \nn\\
&& - 4 g^2\, (X^2 + 2 X^{-1})\, 
        {\hat *\oneone}\,, 
\eea
where  $X = e^{-\frac{1}{\sqrt{6}} \f}$, $\Fh_\2^i = d\Ah_\1^i - \frac{1}{\sqrt{2}}\, g 
\,\e^{ijk}\,\Ah_\1^j \wedge \Ah_\1^k$ and $\Gh_\2 = d\Bh_\1$, and
$\th_{(\a)} = -1$ when $\a = 1$ and $\th_{(\a)}=1$ when $\a=2$.  The Einstein 
summation convention is applied over the indices $i$ and $\a$. The Lagrangian 
(\ref{ds5lag}) can be obtained by applying the following analytic 
continuation
\be \label{adstods}
 g \rightarrow \im g, \quad  \Ah_\2^1 \rightarrow \im \Ah_\2^1, \quad 
 \Ah_\1^i \rightarrow \im\,\Ah_\1^i
\ee
to the gauged $AdS_5$ supergravity Lagrangian \cite{pope1}. 
Compared to the bosonic Lagrangian of $AdS_5$ supergravity, there are 
opposite  
signs in the kinetic terms of $\Ah_\1^i$ fields, the interaction terms of 
$\Ah_\2^1$ to itself, and the term with the coupling constant $g^2$. The term 
$\frac{1}{\sqrt{2}}\, g \,\e^{ijk}\,\Ah_\1^j \wedge \Ah_\1^k$ in $\Fh_\2^i$ also 
has an opposite sign compared to the AdS case. For convenience, 
we have set the Newton constant $G=1$. 

 The Lagrangian (\ref{ds5lag}) gives rise to the following equations of motion 
\bea \label{eom}
  d(X^{-1} \hat * dX) &=& \frac{1}{3} X^4 \hat *\Gh_\2 \wedge \Gh_\2 
                          -\frac{1}{6} X^{-2}(-\hat * \Fh_\2^i \wedge
                          \Fh_\2^i + \th_{(\a)} \hat * \Ah^\a_\2 \wedge \Ah^\a_\2)
                         + \frac{4}{3}\,g^2(X^2-X^{-1})\hat * \oneone, \nn \\
  d(X^4 \hat * \Gh_\2) &=& -\hlf \th_{(\a)} \Ah^\a_\2 \wedge \Ah^\a_\2 
                           + \hlf \, \Fh_\2^i \wedge \Fh_\2^i, \nn \\
  d(X^{-2} \hat * \Fh_\2^i) &=& \sqrt{2}\,g X^{-2} \e^{ijk}\,\hat * 
\Fh_\2^j \wedge \Ah_\1^k - 
                              \Fh_\2^i \wedge \Gh_\2, \nn \\
  X^2 \hat * \Fh^\a_\3 &=& g\,\Ah^\a_\2 \, , \nn \\
  \Rh_{\sub{MN}} &=& 3 X^{-2} \partial_{\sub{M}}X\,\partial_{\sub{N}}X + 
                     \frac{4}{3}\,g^2 (X^2 + 2 X^{-1})\gh_{\sub{MN}}
                    + \hlf X^4 \left[\Gh_{\sub{M}}{}^{\sub{P}} \Gh_{\sub{NP}} 
                                  - \frac{1}{6}\gh_{\sub{MN}}\,(\Gh_\2)^2\right] \nn \\
                 & & + \hlf X^{-2} \left[-\Fh^i_{\sub{M}}{}^{\sub{P}} \Fh^i_{\sub{NP}} 
                     + \frac{1}{6} \gh_{\sub{MN}}\,(\Fh^i_\2)^2\right]
                     + \hlf X^{-2} \th_{(\a)}\, 
                     \left[ \Ah^\a_{\sub{M}}{}^{\sub{P}} \Ah^\a_{\sub{NP}} - 
                     \frac{1}{6} \gh_{\sub{MN}}\,(\Ah^\a_\2)^2\right] ,
\eea 
where $\Fh^\a_\3 = d\Ah^\a_\2 + g\,\Bh_\1 \wedge \Ah^\a_\2$. 
To 
have a consistent reduction, we take $g=k$ and the following reduction ansatz
\bea
  {d\sh_5}^2 &=& dz^2 + \cos^2(kz)\:{ds_4}^2,  \label{ansz1}\\
  \Ah^1_\2 &=& -\frac{1}{\sqrt{2}} \cos(kz)\,*\!F_\2, \quad
       \Ah^2_\2 = - \frac{1}{\sqrt{2}} \sin(kz)\,F_\2,  \label{ansz2}\\
  \Ah^1_\1 &=& \frac{1}{\sqrt{2}}A_\1, \label{ansz3}
\eea
where the four dimensional fields have Minkowskian signature. All other fields,
$\Ah_\1^2,\Ah_\1^3, \Bh_\1,\phi$,  
are set to zero. 
The ansatz (\ref{ansz2}) and (\ref{ansz3}) imply the photon field 
in 4 dimensions is derived from the two-form fields 
$\Ah_\2$ in 5-dimensional de Sitter supergravity, which is different from the 
conventional Kaluza-Klein theory. The equations of the five dimensional
de Sitter supergravity give the following four dimensional equations 
\bea 
   R_{\m\n} &=&  - \hlf ( {F_\m}^\l F_{\n\l}
 -\frac{1}{4} F_\2^2   g_{\m\n} ) + 3 k^2 g_{\m\n}, \label{EM-sys1}\\
d(*F_\2) &=& 0. \label{EM-sys2}
\eea
These are nothing but the  bosonic equations of motion for gauged
$\cN=2$ de Sitter supergravity in four dimensions.
Unlike the usual Einstein-Maxwell equations, the Maxwell term 
in \eq{EM-sys1} takes on an opposite sign which is
characteristic of the de Sitter supergravity. 
The reduction for the
fermionic fields goes through similarly and one obtains the full $\cN=2$ de 
Sitter supergravity in four dimensions. 
The solution of \eq{EM-sys1}, \eq{EM-sys2}  preserves  \cite{pope2} half of the  
five dimensional de Sitter supersymmetries.

\section{Black ring solution with a positive cosmological constant}

The following
``charged de Sitter C-metric''  is a solution to 
the equations \eq{EM-sys1} and \eq{EM-sys2}:
\footnote{One can also take $A_t=q y$ or $A_y =q t$.}
\bea   
&&  {ds_4}^2 = \frac{1}{A^2(x-y)^2} 
          \left[ G(y)\,d t^2 - \frac{dy^2}{G(y)}+\frac{dx^2}{\Gt(x)}
          + \Gt(x) d\varphi^2 \right] , \label{ds41}\\
&& A_\varphi = q x +c_0, \label{ds42} 
\eea
where the coefficient functions $G(\xi)$ and $\Gt(\xi)$ are quartic,
\bea   \label{G0a}
  G(\xi) &=&  q^2 A^2 \xi^4 + a_3 \xi^3 + a_2 \xi^2 + a_1 \xi + a_0,  \nn \\
  \Gt(\xi) &=& G(\xi)-k^2/A^2. 
\eea
Here the 4-dimensional cosmological constant is $\L = 3k^2 > 0$ and the
constants $c_0$, $a_{0,1,2,3}$ and
$A>0$ are arbitrary. The metric \eq{ds41} is a C-metric with a
positive cosmological constant. It 
describes  a pair of uniformly accelerated black holes
in a spacetime with a positive cosmological constant 
\footnote{
The authors of \cite{dsc} claim that the spacetime is asymptotically de
Sitter. We will demonstrate explicitly below, this is not true. See
footnote \ref{foot} below.
}, and $A$ is the uniform 
acceleration of the black holes \cite{dsc}.

One can construct a   solution in the 5-dimensional de Sitter
supergravity theory by
substituting \eq{ds41} into (\ref{ansz1}), 
\be  \label{dsbr}
  {d\sh_5}^2 =  dz ^2 + \frac{\cos^2(kz)}{A^2(x-y)^2} 
          \left[
G(y)\,d t^2 - \frac{dy^2}{G(y)} +\frac{dx^2}{\Gt(x)} + \Gt(x) d\varphi^2 
\right] \, .
\ee  
By construction, our solution preserves half of the supersymmetries of the 
5-dimensional de Sitter supergravity.

In the following we will choose the functions $G, \Gt$ to be even. 
Moreover we consider the case that all the roots 
of $G$ and $\Gt$ are real and distinct, and specify the roots as 
$\xi_1 < \xi_2 < \xi_3 <\xi_4$ and $\xit_1 < \xit_2 < \xit_3 <
\xit_4$. 
It is easy to show that, by rescaling the coordinates appropriately, one
can always choose $a_0=1=-a_2$. Therefore we consider $G$, $\Gt$ of the form
\bea \label{newG0a}
G(y)  &=& 1 -y^2 + q^2 A^2 y^4 =  q^2 A^2 (y^2 - \xi_3^2)(y^2-\xi_4^2),\nn\\
\Gt(x)   &=& 1 -k^2/A^2 - x^2 + q^2 A^2 x^4 = q^2 A^2 (x^2 -
\xit_3^2)(x^2-\xit_4^2).
\eea
The existence of the roots $\xit_2,\xit_3$ imposes that 
\be \label{Gk}
k^2/A^2 < 1. 
\ee 
Explicitly the roots are given by
\be  \label{roots}
\xi_{3,4}^2 = \frac{1}{2q^2A^2}(1\mp\sqrt{1-4 q^2 A^2} \; ), \qquad
\xit_{3,4}^2 = \frac{1}{2q^2A^2}(1\mp\sqrt{1-4 q^2 A^2(1- k^2/A^2)}
\; ),
\ee
where the minus (or plus) sign  corresponds to $\xi_3, \xit_3$ (or
$\xi_4,\xit_4$).  And $q$ is in the range
\be
0 \leq q \leq \frac{1}{2A}.
\ee

When $q=0$, $G(y) =1-y^2$, and 
the metric \eq{ds41} describes a particular C-metric with a
cosmological constant. In the terminology of \cite{dsc}, it is called the
massless uncharged de Sitter C-metric. 
We note that although the metric 
satisfies the maximal symmetric space
condition 
\be
K_{\m\n\l\r}:= R_{\m\n\l\r} - k^2( g_{\m\l}g_{\n\r} -g_{\m\r}
g_{\n\l}) =0
\ee 
when $q=0$ 
\footnote{ \label{foot}
By examining the behaviour of the curvature invariants similar to that 
of \eq{invar} below,  
\cite{dsc} claims that the metric \eq{ds41} approaches the 4-dimensional 
de Sitter space asymptotically 
as $r \to \infty$. However this statement is wrong. In fact
the tensor $K_{\m\n\l\r}$
is nonvanishing when $q\neq 0$. For example, it is
$K_{x\varphi x \varphi} = q^2 (4 A x r +1) +a_3 r/A$. For the
massless-uncharged C-metric, $G=1-y^2$ and one can check that
$K_{\m\n\l\r} =0$ identically. 
},
it is not the same as de Sitter space as there is an accelerating horizon
besides the cosmic horizon 
\footnote{The
situation is similar to the difference between the Minkowskian space and
the Rindler space: locally they are the same, but the Rindler space has
an acceleration horizon and a temperature parametrized by the
acceleration parameter. } .
Only when $A=0$ also, \eq{ds41} reduces to a pure de Sitter spacetime
under a suitable coordinate transformation \cite{dsc}.
The solution is 
parametrized by $k$ which characterizes the cosmological constant 
and the acceleration parameter $A$. 
The corresponding 
five-dimensional metric \eq{dsbr} is locally de Sitter and 
can be thought of as the background upon which a general 
solution with $q\neq 0$ is obtained by turning on the the parameter 
$q$.

\begin{figure}
\label{fig1}
\psfrag{x}{$x$} 
\psfrag{y}{$y \geq \xit_3$} 
\psfrag{xi1}{$\xi_1$}
\psfrag{xi2}{$\xi_2$}
\psfrag{xi3}{$\xi_3$}
\psfrag{xi4}{$\xi_4$}
\psfrag{xit1}{$\xit_1$}
\psfrag{xit2}{$\xit_2$}
\psfrag{xit3}{$\xit_3$}
\psfrag{xit4}{$\xit_4$}
\psfrag{G}{$G$}
\psfrag{Gt}{$\Gt = G-k^2/A^2$}
\begin{center}
{\scalebox{1}{ \includegraphics{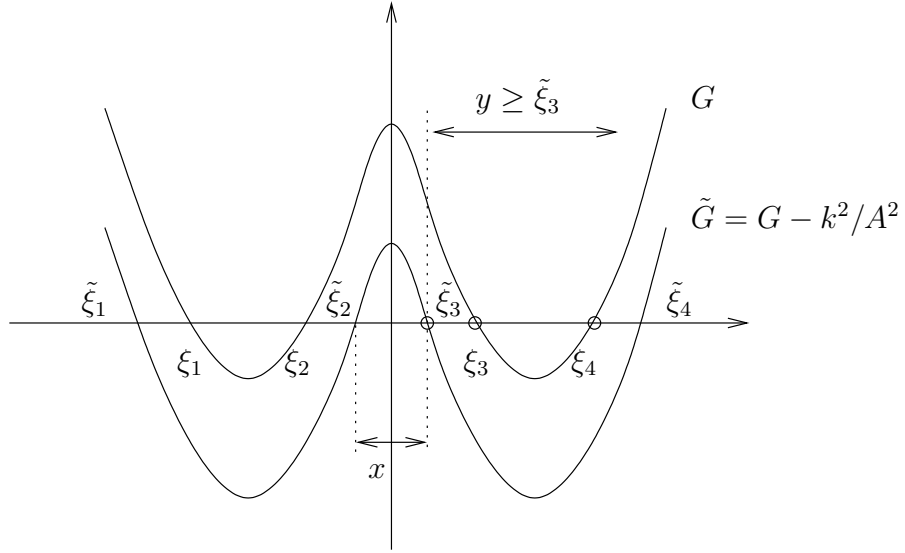}} }
\caption{Choice for the range of coordinates}
\end{center}
\end{figure}

In the follwing, we will consider the solution 
\eq{dsbr}   with $q\neq 0$, $G,\Gt$ given by
\eq{newG0a} and constraint $k^2/A^2<1$.
We will show it describes a solution  with a 
cosmological constant $\hat{\L} = 6 k^2$ and 
admits horizons of ring topology $S^1\times S^2$.
The solution is parametrized by the parameters $k,q,A$. 
We will see later that a non-zero $q$ corresponds to a non-zero dipole
charge for our solution.

To demonstrate our claim,  we need to   specify an appropriate 
range of $x,y$. 
We choose $x$ to be $\xit_2 \leq x \leq \xit_3$ so that $\Gt(x)\geq 0$,
and $ \xit_3 \leq y < \infty$. See the figure.  
In 5 dimensions, for the constant $(z, t, y)$ slices of (\ref{dsbr}), 
conical  singularities appear at $x=\xit_2$ and $x=\xit_3$ 
as $g_{\varphi\varphi}=0$ at these points. 
To avoid conical singularities we identify $\varphi$ with the period
\be  \label{period}
  \D\varphi \,\Big\vert_{x=\xit_2} = \frac{4\pi}{|G'(\xit_2)|},   \qquad
  \D\varphi \,\Big\vert_{x=\xit_3} = \frac{4\pi}{|G'(\xit_3)|}.
\ee
Since we have chosen $G$ to be even, the two periods are automatically
equal and we have
\be \label{period-phi}
\D \varphi = \frac{2 \pi}{q^2 A^2 |\xit_3 (\xit_4^2-\xit_3^2)|}.
\ee 
With this period, 
the constant $(z, t, y)$ surface has an 
$S^2$ topology spanned by $x$ and $\varphi$, 
with the north and south poles at $x=\xit_2$ and $x=\xit_3=-\xit_2$.
Due to the warp factor $\cos^2(k z)$ in the metric, the coordinate
$z$ is periodic 
$-\pi/2k < z \leq \pi/2k$
with period
\be
\D z = \frac{ \pi}{k}.
\ee 
Therefore the sections at constant $t$, $y$  has a ring topology 
$S^1 \times S^2$. It should also be clear now why a ring topology of
horizon is possible only for the case (iv) as  the warp factor in all the 
other cases is not a periodic function, implying the warp coordinate
is non-periodic. But our construction can give new black tube solution with
horizon of $S^2\times R$ topology for all cases of the brane world reduction.
The interested readers can fill out the details easily. See also
\cite{brihaye} for some black string solutions with positive cosmological
constant.

Next we examine the asymptotic behaviour of the metric as $x \to y \to
\xit_3$. It is convenient 
to apply the coordinate transformation \cite{dsc}
\be 
0< r=\frac{1}{A(y-x)} <\infty
\ee 
to  (\ref{dsbr}). For the more general form of $G$ and $\Gt$ as in
\eq{G0a}, the curvature invariants are (for general $x, y ,z$):
\bea \label{invar} 
\Rh &=& 20k^2,  \nn \\
\Rh_{MN}\Rh^{MN} &=& 80k^4 + \frac{4 q^4}{\cos^4(kz) r^8}, \\
\Rh_{MNPQ}\Rh^{MNPQ} &=& 40k^4 + 
\frac{192 q^4 A^2 x^2 + 96 q^2 a_3 x + 12 a_3^2/A^2}{\cos^4(kz) r^6}
\nn \\
&& + \frac{192 q^4 A x + 48 q^2 a_3 / A}{\cos^4(kz) r^7}
+ \frac{56 q^4}{\cos^4(kz) r^8}.        \nn  
\eea 
As $r \rightarrow \infty$, the terms inverse proportional to higher order $r$ vanish in 
(\ref{invar}), and the curvature invariants approach those of $dS_5$ spaces, 
$\Rh \rightarrow 20k^2$, $\Rh_{MN}\Rh^{MN} \rightarrow 80k^4$, $\Rh_{MNPQ}\Rh^{MNPQ}
\rightarrow 40k^4$. It is suggestive that the metric may approaches  
$dS_5$ as $r \to \infty$. To check this, we should check whether the tensor
\be \label{max-symm}
\hat{K}_{\m\n\l\r}:= \Rh_{\m\n\l\r} - 
k^2( \gh_{\m\l}\gh_{\n\r} -\gh_{\m\r} \gh_{\n\l})
\ee
approaches zero or not. One can easily check that it is not as long
as $q \neq 0$ 
For example, $\hat{K}_{x \varphi x \varphi} = \cos(k z)^2 (4 A q^2 x r + a_3 r/A+q^2)$.
Thus our metric \eq{dsbr} does not approach $dS_5$. However, it 
is 
\bea \label{ein}
 \frac{\Rh_{zz}}{\gh_{zz}} &=& 4 k^2, \nn \\
 \frac{\Rh_{tt}}{\gh_{tt}} &=& \frac{\Rh_{yy}}{\gh_{yy}} = 4 k^2 + 
\frac{q^2}{r^4 \cos^2(kz)}, \\
 \frac{\Rh_{xx}}{\gh_{xx}} &=& \frac{\Rh_{\varphi\varphi}}{\gh_{\varphi\varphi}} 
= 4 k^2 - \frac{q^2}{r^4 \cos^2(kz)}. \nn
\eea  
Therefore as $r\to \infty$, the metric satisfies the   Einstein equation with
a cosmological constant $\hat{\L} = 6 k^2$ and we have a 
solution  with a positive cosmological constant.

Our metric \eq{dsbr} has the Killing vectors $\del/\del t$  and $\del/\del \varphi$. 
In the region $\xi_3 \leq y \leq \xi_4$,
$G(y)<0$, $t$ is timelike and $y$ is spacelike. At the endpoints
$y=\xi_3$ or $\xi_4$, the coordinates break down. Let us introduce new coordinate 
$v$ by  
\be
dv = dt + \frac{dy}{G(y)}. 
\ee
In the new coordinates, the metric taking the form
\be  \label{dsbr1}
  {d\sh_5}^2 =  dz ^2 + \frac{\cos^2(kz)}{A^2(x-y)^2} 
          \left[
G(y)\,d v^2 - 2 dv dy  +\frac{dx^2}{\Gt(x)} + \Gt(x) d\varphi^2 
\right] \, 
\ee  
is regular. 
It is easy to show that the surface $y= y_0$ 
(where $y_0 =\xi_3$ or $\xi_4$)
is a Killing horizon of the Killing vector field
\footnote{We will justify the normalization of the normalization of the Killing
vector \eq{killing-v} later. } 
\be \label{killing-v}
\eta= \frac{\del}{\del v}. 
\ee 
The horizon has $S^2\times S^1$ topology and 
has the surface gravity
\be \label{kappa}
\k = \frac{ |G'(y_0)|}{2} =  q^2 A^2 y_0 (\xi_4^2-\xi_3^2)
\ee
and the horizon area 
\be \label{area}
\cA = \frac{2 \pi^2 }{k q^2 A^4 }\cdot \frac{1}{ (y_0^2 -
\xit_3^2)  (\xit_4^2-\xit_3^2) }
\ee 
for $y_0 =\xi_3$ or $\xi_4$.
In terms of $q$, we have
\be
\k_\ri = \frac{\sqrt{1-4 q^2 A^2}\sqrt{1\mp \sqrt{1-4 q^2 A^2}}}{\sqrt{2} q A} 
\ee
and
\be
\cA_\ri  = 
\frac{ 4 \pi^2 q^2}{k}\frac{1}{ 
\sqrt{1-4 q^2 A^2 (1-k^2/A^2)} \mp \sqrt{1-4 q^2 A^2}}
\frac{1}{\sqrt{1-4 q^2 A^2 (1-k^2/A^2)}} 
\ee
where  sign $\ri = -$ (resp. $+$) is for $y_0 = \xi_3$ (resp. $y_0 = \xi_4$).
Note that
\be
\cA_{-} > \cA_{+} >0. 
\ee
The horizon at $y= \xi_3$ is the cosmological horizon. It is 
at a larger $r$ and has a larger area. As one decreases $y$
(i.e. increases $r$) to the range $\xit_3 \leq y < \xi_3$,
$G(y)$ becomes positive, $y$ becomes timelike and $t$ becomes
spacelike. The situation is similar to the de Sitter space where the
timelike Killing vector becomes spacelike  as one
goes outside the cosmological horizon.
On the other hand, if one increases $y$ (i.e. decreases $r$)  
until $\xi_4$, we reach the 
the black ring  event horizon. 
The black ring horizon is at a smaller $r$ and has a smaller
area. 
The metric can be continued
beyond the black ring horizon to $y>\xi_4$, until $y=\infty$
(correspondingly $r=0$), which is a curvature singularity. 
 
As is clear from \eq{invar}, the horizons are singular at $z=
\pi/2k$. In fact the $S^2$ factor of the horizon 
shrinks to zero size there. However the solution is unstable
near the tip due to the Gregory-Laflamme instability \cite{GL}. 
Two different possible final states has been suggested in the
literature, different in  whether the horizon will pinch off \cite{GL} 
or not \cite{HM}. In our case, if the horizon pinches off at the tip,
then a black hole solution with a regular horizon of 
topology $S^3$ will be formed. On the other hand, 
if the horizon does not pinch off, 
then a new horizon is expected to form to surround the neck region
and a black ring with a regular horizon of topology
$S^1\times S^2$ will be formed. 
We tend to believe the Horowitz-Maeda sceranio  is more likely for our case. 
Moreover, if we
allow rotation in $\varphi$, then, at least for sufficiently large 
angular momentum,   the second sceranio appears to be
more favorable. We conjecture this is the case. 
If this is true, our solution with a singular ring 
horizon can be intrepretated as evidence that a black ring in de
Sitter space do exists.

Despite the singularity, the
surface gravities and  the areas are finite and one may consider the
thermodynamics associated with the  horizon.
The  horizons have the Hawking temperatures 
\be \label{hawkingT}
T_\ri = \frac{\k_{\ri}}{2 \pi} = 
\frac{\sqrt{1-4 q^2 A^2}\sqrt{1\mp \sqrt{1-4 q^2 A^2}}}{2 \sqrt{2} \pi
  q A} . 
\ee 
The same
temperature can also be obtained by Wick rotating the metric \eq{dsbr}
with $t = i \tau$. The $y$-$\t$ part of the Euclidean metric has conical
singularities at $y_0$ unless $\t$ is periodic with period
\be
\D \tau = \frac{4 \pi}{|G'(y_0)|} = \frac{2 \pi}{\k}.
\ee 
This gives immediately the temperature \eq{hawkingT}. 
This also justify  the choice of the
normalization of the Killing vector \eq{killing-v}. 
 
As usual, the horizons carry a Bekenstein-Hawking entropy
\be \label{S}
S_{\ri}= \frac{\cA_{\ri}}{4} = 
\frac{ \pi^2 q^2}{k}\frac{1}{ 
\sqrt{1-4 q^2 A^2 (1-k^2/A^2)} \mp \sqrt{1-4 q^2 A^2}}
\frac{1}{\sqrt{1-4 q^2 A^2 (1-k^2/A^2)}} 
\ee
where the sign $-(+)$ is for the cosmological (black ring) horizon. 
For small $q$, the entropy for 
the cosmological horizon $S_c$ behaves as
\be
S_c = \frac{\pi^2}{2 k^3} -\frac{\pi^2}{2k} q^2 
-\frac{\pi^2}{2k}(4A^2-3k^2) q^4 + O(q^6),
\ee
and is a decreasing function of $q$.
In fact it is easy to verify that
\be
S_c \le \frac{\pi^2}{2 k^3} := S_{\rm de \; Sitter},
\ee
where $S_{\rm de \; Sitter}$ is the entropy for 5-dimensional pure de
Sitter space of the same cosmological constant. This reminds us of the 
N-bound  proposal \cite{bousso} of Bousso which states that
the total entropy of a spacetime with a cosmological constant is bounded
by the  entropy of the pure de Sitter space of the same cosmological
constant. In our case, the total entropy $S_T$ is 
the sum of the entropy 
of the black ring  and the entropy  of the cosmological horizon
\footnote{The entropy is defined by  
$S = \beta (\partial /\partial \beta -1) I$, where
$I$ is the action of the Euclidean solution.
As in the Schwarzschild-de Sitter black
hole, both the black ring (or the black hole) horizon and the cosmological 
horizon appear in the real Euclidean geometry and so
they both contribute to the entropy of our solution.  
We thank Simon Ross for explanation of this point. 
}. 
Quite remarkably, we obtain the result
\be
S_T = S_{\rm de \; Sitter}.
\ee
Thus the N-bound is  precisely satuated. 
This is amazing especially because there are matters 
violating the usual form of energy condition due to their
negative kinetic terms in the dS supergravity; and as a result one may
expect the N-bound to be violated. 
We view this as an evidence that there is a sensible quantum gravity
description of the singular black ring.

We also note that the specific heat  
\be
C_{\rm i}:= T \frac{\del S_{\rm i}}{\del T} = 
\frac{\pi^2 q^2}{k} \frac{\sqrt{1-4 q^2 A^2}\left(1\mp\sqrt{1-4 q^2 A^2}
    \right)}
{\left(1-4 q^2 A^2(1-k^2/A^2)\right)^{3/2} 
\left(1\mp  \sqrt{1 - 4 q^2 A^2} (1+2 q^2 A^2)\right)} ,
\ee
is positive for the cosmological horizon and negative for the black ring
horizon, meaning that the cosmological horizon   is thermally
stable while the black ring horizon is thermally unstable. This is
similar to that of a de Sitter Schwarzschild black hole.

In addition to the metric, our solution is supported by nontrivial
two-form $\Fh^1_{(2)}$ and three-forms $\Fh^\a_{(3)}$ ($\a=1,2)$. They
obey the  equation of motion: $d (\hat{*} \Fh^1_{(2)}) =0$ and 
$\hat{*} \Fh^\a_{(3)} = k \Ah^\a_{(2)}$. The non-standard form of
equation of motion of the three-forms does not lead to any conserved
charge. As in \cite{dipole-ring}, the 
two-form leads to the dipole charge: 
\footnote{The extra factor of $\sqrt{2}$ here and in \eq{pot} below
is due to the nonstandard  normalization of fields in the Lagrangian \eq{ds5lag}.
}
\be
q_e = \frac{\sqrt{2}}{4 \pi} \int_{S^2} \hat{*} \Hh, 
\ee
where the integral is carried over any $S^2$ which can be deformed to
an $S^2$ on the cosmological horizon 
\footnote{
The inner black ring horizon carries exactly the same dipole charge.
However as we will be interested in the thermodynamics of the
cosmological horizon only, it will not be relevant for our
consideration.
}. Here 
$\Hh$ is the dual field strength 
\be
\Hh = d \Bh = \hat{*}\; \Fh^1_{(2)}.
\ee
The dual 2-form potential is
\be \label{Btz}
\sqrt{2} \Bh_{tz} =q y + c_1, 
\ee
where $c_1$ is a constant.
$q_e$ is well defined due to the equation of motion $d(\hat{*} \Hh) =0$. 
For our solution, $\Fh^1_{x \varphi} =q/\sqrt{2}$ and 
the dipole charge is 
\be \label{dipole-charge}
q_e = - \frac{q}{\sqrt{1-4 q^2 A^2 (1-k^2/A^2)}}.
\ee

Next we would like to determine the mass of our solution. 
Generally to determine the mass associated with a 
given gravitational configuration, 
a well known procedure due to Brown and York is to start with a  
suitably defined quasi local
stress tensor on the boundary of a given region of spacetime \cite{BY}
\be
T^{\m\n} := \frac{2}{\sqrt{-\c}}\frac{\d S_{\rm grav}}{\d \c_{\m\n}}.
\ee
Here $\c_{\m\n}$ is the boundary metric and $S_{\rm grav}$ is the gravitational 
action thought of as a functional of the boundary metric. 
The resulting stress tensor
typically diverges as the boundary is taken to infinity. To obtain a
finite stress tensor, one may try to add an appropriate boundary 
term which  does
not affect the bulk equation of motion
to cancel the divergences.
The original proposal of Brown-York utilises a subtraction derived by
embedding the boundary with the same intrinsic metric $\c_{\m\n}$
in some reference spacetime. However this is not always possible. 
Due to interests  in AdS/CFT and dS/CFT correspondence, the quasi local 
stress tensor for spacetime which is
asymptotically Anti-de Sitter or 
asymptotically de Sitter has been worked out in
\cite{tensor1,tensor2} using a different subtraction procedure. 
The required counter terms are constructed in terms of the  boundary
curvature invariants and are fixed essentially uniquely by the
finiteness requirement of the stress tensor. Unfortunately none of these
methods help with our present case where the asymptotic behaviour of the
metric is rather complicated.
Thus instead of trying to look for a first
principle determination of the mass, 
we will {\it assume} the validity
of the first law of thermodynamics 
and use it to determine the mass of our solution. 
In particular since the mass should be defined from the asymptotic
infinity, i.e. outside of the cosmological horizon, the relevant first
law of thermodynamics shall be the one related to the cosmological
horizon.

The first law of thermodynamics for de Sitter black hole
was first obtained by \cite{GW}, 
and the form presented contains contributions
from the black hole horizon as well as the cosmological horizon. 
It has been noted \cite{cosmo} that
for a wide class of de Sitter black holes, 
the cosmological horizon satisfies a first law of
thermodynamics by itself remarkably. 
And it is natural to guess that  given an 
appropriate definition of the energy, the
cosmological horizon will always satisfy the first law of thermodynamics. 
We will assume so for our case 
and use it to derive the mass of the solution.
For dipole ring,  there is however a new ingredient. 
It was first noted by Emparan \cite{dipole-ring}  that the dipole charge
appears in the first law of thermodynamics in the same manner as a
global charge. This is not expected
from the general derivation of the first law by Sudarsky and Ward 
\cite{SW}  and the explanation has been
given by Copsey and Horowitz \cite{CH}. 
Usually the gauge potential $B$ can be globally defined and
non-singular everywhere outside and on the horizon. However this is
not compatible with the assumptions of a non-vanishing dipole charge
and that $B$ is invariant under spacetime symmetries. 
Since $B$ is defined up to gauge transformation, one can choose
any gauge to try to determine the consequence of this incompatibility.
A particularly transparent gauge is to have  $B_{t\psi}$ as the
only nonzero component and to have $B_{t\psi}$   vanishes at the infinity. 
As demonstrated by \cite{CH}, this implies that  
$B_{t\psi}$  vanishes at the rotation axis (hence violation of the above 
stated conditions) and    
$B_{t \psi}$ necessarily diverges at the horizon. As a result a new dipole term 
arises in the first law. In our case, we also have a dipole charge. For
it to make contribution to the first law, we need to examine the
behaviour of the dipole potential 
\be \label{pot}
\phi_e = - \frac{\sqrt{2} \pi}{2} B_{t\bar{z}}{}\Big|_{\rm horizon},
\ee
where $\bar{z} = (\pi/\Delta z) z = k z$ is the canonical normalized angular
variable and it is evaluated at the cosmological or the black ring
horizon. 
Note that $B_{tz} = \eta^\m B_{\m\n} (\del/\del z)^\n$ and
so we expect that $B_{tz}$ to be vanishing at both horizons. However
this is impossible for ours \eq{Btz}. This means $B$ must be singular
somewhere.
Choosing  the constant $c_1$ such that $\phi_e=0$ at infinity and 
follows the same argument as \cite{CH}, we obtain a
contribution $\phi_e d q_e $ to the first law of thermodynamics for 
the cosmological horizon
\be \label{first-law}
d E =T dS +  \phi_e d q_e .
\ee
Here $q_e$
is the dipole charge \eq{dipole-charge} and $\phi_e$ is the
dipole potential evaluated at the cosmological horizon 
\be
\phi_e = - \frac{\pi q}{2k} (\xi_3 - \xit_3).
\ee 

We note that $dE$ is an exact differential and $E$ is well defined.
In fact although our solution is parametrized by the three parameters 
$k, A$ and $q$, one should think of $k, A$ as specifying the background and 
$q$ as specific to the solution. By
increasing $q$ from 0 to a nonzero value, we get to our 
solution. We have
\be \label{E}
M:= E -E_0= \int_0^{q} (T \frac{d S}{d q} + \phi_e \frac{d q_e}{dq}
) dq ,
\ee
where $E_0$ is the energy of the background (i.e. when $q=0$). 
The mass $M$ is given by the 
difference of energy $ E - E_0$ 
above the background. The first integral 
\be \label{term1}
\int_0^{q} T \frac{d S}{d q} dq = 
- \frac{\pi}{2\sqrt{2}kA} \int_0^q \frac{\sqrt{1-\sqrt{1-4 q^2 A^2}}}{(1-4 q^2
  A^2(1-k^2/A^2))^{3/2}} dq
\ee
is manifestly negative.
For the second integral  
\be
\int_0^{q} \phi_e \frac{d q_e}{dq} dq = 
- \frac{\pi}{2 k} \int_0^q q \xi_3 \frac{d q_e}{dq} dq 
+ \frac{\pi}{2 k} \int_0^q q \xit_3 \frac{d q_e}{dq} dq ,
\ee
it is easy to show that the first piece
\be
- \frac{\pi}{2 k} \int_0^q q \xi_3\frac{d q_e}{dq} dq= 
\frac{\pi}{2\sqrt{2}kA} \int_0^q \frac{\sqrt{1-\sqrt{1-4 q^2 A^2}} \;dq}
{(1-4 q^2A^2(1-k^2/A^2))^{3/2}}  = - \int_0^{q} T \frac{d S}{d q} dq
\ee
cancels exactly the entropy contribution \eq{term1} and so
\bea
M &=&  \frac{\pi}{2 k} \int_0^q q \xit_3 \frac{dq_e}{dq} dq \nn\\
&=&
-\frac{\pi}{4 \sqrt{2} k A^2 \sqrt{1-k^2/A^2}} 
\left(
\frac{\sqrt{1+Y}- \sqrt{2} Y}{Y} 
- \frac{1}{2} \ln (\frac{\sqrt{1+Y} +1}{\sqrt{1+Y}
-1}\frac{\sqrt{2}-1}{\sqrt{2}+1})
\right), 
\eea 
where
\be
Y:=\sqrt{1-4q^2A^2(1-k^2/A^2)} = |q/q_e| \leq 1 .
\ee

We note that $M$ is negative, showing that 
our solution has a smaller energy 
with respect to the background's. This is similar to
the case of black holes in asymptotically de Sitter space where the pure
de Sitter spaces is always more massive than the Schwarzschild-de Sitter
black holes in the corresponding dimensions. This has leaded
\cite{tensor2} to the conjecture that any asymptotically dS space whose
mass exceeds that of pure dS space must contain a cosmological
singularity. The fact that our solution carries a negative mass in a
spacetime with a positive cosmological constant leans support to the 
following {\it generalized maximal mass conjecture}: 
In any spacetime with a positive
cosmological constant, if the addition of matter leads to an increase 
in energy, the 
resulting solution must contain a cosmological singularity.

We remark, however, that this cannot be the complete statement and presumably
more specific conditions needed to be specified 
\footnote{Explicit examples are known which violate the N-bound proposal
\cite{vio1,vio2} as well as the maximal mass conjecture \cite{vio2}.
Presumably this is due to the fact that certain asymptotic energy
condition is not satisfied. To our knowledge, the precise condition for
the N-bound proposal or the maximal mass conjecture to hold has not been
formulated explicitly. We thank Simon Ross for discussions on this
matter.} .  
Our  solution provides an explicit example which
satisfies both the N-bound proposal and the maximal mass conjecture, and
may help one to identify the appropriate conditions.

\section{Discussions}

In this paper we have constructed a singular  black ring solution
in the 5-dimensional de Sitter supergravity theory. The solution has
singular horizons whose singularity occurs when the warp factor vanishes. 
Despite the singularity in the horizon, thermodynamic quantities such
as temperature and entropy are well defined and finite. 
It is quite amazing to find that the entropy and mass of our solution 
are  consistent with what one would expect for solution in a
spacetime with a positive cosmological constant.  Moreover the
N-bound is precisely satuated for our solution.
It will be interesting to
understand its physical significance.

Our construction relies on the brane-world reduction ansatz.
In the case of a negative cosmological constant, 
this ansatz  does not give any solution with ring topology, not even
singular one.
This can be easily understood as our metric ansatz does not include
rotation, while we expect a black ring in a spacetime with negative
cosmological constant to be rotating 
since there is no repulsive force from the background. 
It maybe possible to generalize our
ansatz to include rotation and use 
this  to construct  a black ring with a
negative cosmological constant.

Our solution has a metric which is not asymptotically de Sitter and so
one cannot apply the usual subtraction procedure to the
Brown-York stress tensor. We remark that in case the deviation
from asymptotically (anti-)de Sitter behaviour is caused by a
nontrivial dilaton potential, a well defined boundary stress tensor
can be constructed \cite{cai-ohta}. Our solution is deformed
from the asymptotically de Sitter behaviour by the
presence of nontrivial form fields in the action. 
It may be possible to devise a similar procedure 
to construct a well defined boundary stress tensor and to obtain the
mass. With this one can verify the first law of thermodynamics \eq{first-law}
independently. It is also interesting to derive the first law of
thermodynamics by generalizing and extending the
procedure of \cite{SW} to include a
positive cosmological constant.

De Sitter supergravity
contains matter fields with wrong sign kinetic terms, 
one may therefore worries about
the stability of our solution. 
As argued in \cite{hull0}, it is possible that the ghost
modes are artifacts purely because of a truncation to the supergravity
limit and are not present in
the full string theory. Therefore one can
expect that  this unpleasant feature of the de Sitter supergravity will be go
away once all the string modes and string corrections are included; 
and the black ring will
be a solution to the full string equations of motion which are manifestly
free of any ghost modes. In particular since
our solution is supersymmetric, we do expect our solution to be stable, at least
classically. 
It has been argued \cite{cho-nam} that ghosts fields for some solutions
in de Sitter supergravity does not lead to instability. A similar analysis 
may be performed which support our
guess.

Incidently a very recent preprint \cite{adsbr} 
appears which deals with the existence of black ring 
in Anti-de Sitter space. These authors find that the most general
supersymmetric, asymptotically AdS black hole that admits two rotation
symmetries may have a horizon of topology $S^1 \times S^2$. However 
there is always a conical singularity presents in 
the $S^2$ factor, thus ruling out the possibility of such black ring. 
The authors also suggest that
nonsupersymmetric black ring may exists by increasing the angular
momentum of the singular solution they found. 
In our case, our solution is 1/2 BPS and has one rotation symmetry.
Since a positive 
cosmological constant is repulsive, it should be possible for a 
 de Sitter black ring to exist without any
rotation and we conjecture this is the case.  
It is possible one has to break more supersymmetries. 
It remains a challenging problem to construct explicitly 
a regular black ring with either positive or negative 
cosmological constant.

\vspace{2cm}
\bigskip

{\bf Acknowledgements}

We would like to thank Mohsen Alishahiha, Roberto Emparan, Ruth Gregory,
Veronika Hubeny, 
Takeo Inami,  Clifford Johnson, 
James Lucieti, Mukund Rangamani, Harvey Reall and 
especially Simon Ross for helpful
discussions. We are particularly grateful to Roberto Emparan and Simon Ross 
for reading the revised
version of the manuscript and for giving us numerous valuable 
suggestions and comments.
CSC thanks the theory group of CERN for hospitality where
part of the work was carried out. CSC acknowledges the support of a
PPARC rolling grant and an EPSRC advanced fellowship.
SHD acknowledges support from the Minstry of Education  of Taiwan for 
a research student fellowship.

\bigskip



\begin{thebibliography}{99}



\bm{br}
R.~Emparan and H.~S.~Reall,
  ``A rotating black ring in five dimensions,''
  Phys.\ Rev.\ Lett.\  {\bf 88}, 101101 (2002)
  [arXiv:hep-th/0110260].

\bm{cr} 
  H.~Elvang,
  ``A charged rotating black ring,''
  Phys.\ Rev.\ D {\bf 68}, 124016 (2003)
  [arXiv:hep-th/0305247].
\\
  H.~Elvang and R.~Emparan,
  ``Black rings, supertubes, and a stringy resolution of black hole
  non-uniqueness,''
  JHEP {\bf 0311}, 035 (2003)
  [arXiv:hep-th/0310008].


\bibitem{susy-ring}
H.~Elvang, R.~Emparan, D.~Mateos and H.~S.~Reall,
  ``A supersymmetric black ring,''
  Phys.\ Rev.\ Lett.\  {\bf 93} (2004) 211302
  [arXiv:hep-th/0407065].
\\
I.~Bena and N.~P.~Warner,
  ``One ring to rule them all ... and in the darkness bind them?,''
  arXiv:hep-th/0408106.
\\
H.~Elvang, R.~Emparan, D.~Mateos and H.~S.~Reall,
  ``Supersymmetric black rings and three-charge supertubes,''
  Phys.\ Rev.\ D {\bf 71} (2005) 024033
  [arXiv:hep-th/0408120].
\\
J.~P.~Gauntlett and J.~B.~Gutowski,
  ``General concentric black rings,''
  Phys.\ Rev.\ D {\bf 71} (2005) 045002
  [arXiv:hep-th/0408122].

\bm{minimal}
J.~P.~Gauntlett, J.~B.~Gutowski, C.~M.~Hull, S.~Pakis and H.~S.~Reall,
  ``All supersymmetric solutions of minimal supergravity in five dimensions,''
  Class.\ Quant.\ Grav.\  {\bf 20} (2003) 4587
  [arXiv:hep-th/0209114].

\bm{dipole-ring}
R.~Emparan,
  ``Rotating circular strings, and infinite non-uniqueness of black rings,''
  JHEP {\bf 0403} (2004) 064
  [arXiv:hep-th/0402149].

\bm{CH}
K.~Copsey and G.~T.~Horowitz,
  ``The role of dipole charges in black hole thermodynamics,''
  Phys.\ Rev.\ D {\bf 73} (2006) 024015
  [arXiv:hep-th/0505278].

\bm{radu}
D.~Astefanesei and E.~Radu,
  ``Quasilocal formalism and black ring thermodynamics,''
  Phys.\ Rev.\ D {\bf 73} (2006) 044014
  [arXiv:hep-th/0509144].

\bm{rogatko}
M.~Rogatko,
``Black rings and the physical process version of the first law of
  thermodynamics,''
  Phys.\ Rev.\ D {\bf 72} (2005) 074008
  [Erratum-ibid.\ D {\bf 72} (2005) 089901]
  [arXiv:hep-th/0509150];
  ``First law of black rings thermodynamics in higher dimensional dilaton
  gravity with p+1 strength forms,''
  Phys.\ Rev.\ D {\bf 73} (2006) 024022
  [arXiv:hep-th/0601055];
  ``First Law of Black Rings Thermodynamics in Higher Dimensional Chern-Simons
  Gravity,''
  arXiv:hep-th/0611260.

\bm{entropy}
 M.~Cyrier, M.~Guica, D.~Mateos and A.~Strominger,
  ``Microscopic entropy of the black ring,''
  Phys.\ Rev.\ Lett.\  {\bf 94} (2005) 191601
  [arXiv:hep-th/0411187].
\\
F.~Larsen,
  ``Entropy of thermally excited black rings,''
  JHEP {\bf 0510} (2005) 100
  [arXiv:hep-th/0505152].
\\
S.~Giusto, S.~D.~Mathur and Y.~K.~Srivastava,
  ``A microstate for the 3-charge black ring,''
  arXiv:hep-th/0601193.


\bm{int}
S.~S.~Yazadjiev,
  ``Completely integrable sector in 5D Einstein-Maxwell gravity and derivation
  of the dipole black ring solutions,''
  Phys.\ Rev.\ D {\bf 73} (2006) 104007
  [arXiv:hep-th/0602116].
\\
H.~Iguchi and T.~Mishima,
  ``Solitonic generation of five-dimensional black ring solution,''
  Phys.\ Rev.\ D {\bf 73} (2006) 121501
  [arXiv:hep-th/0604050].
\\
S.~Tomizawa and M.~Nozawa,
  ``Vaccum solutions of five-dimensional Einstein equations generated by
  inverse scattering method. II: Production of black ring solution,''
  Phys.\ Rev.\ D {\bf 73} (2006) 124034
  [arXiv:hep-th/0604067].


\bm{review}
  R.~Emparan and H.~S.~Reall,
  ``Black rings,''
  Class.\ Quant.\ Grav.\  {\bf 23} (2006) R169
  [arXiv:hep-th/0608012].


\bibitem{bena}
  I.~Bena and P.~Kraus,
  ``Microscopic description of black rings in AdS/CFT,''
  JHEP {\bf 0412} (2004) 070
  [arXiv:hep-th/0408186].

\bm{dowker}
  F.~Dowker, J.~P.~Gauntlett, D.~A.~Kastor and J.~H.~Traschen,
  ``Pair creation of dilaton black holes,''
  Phys.\ Rev.\ D {\bf 49} (1994) 2909
  [arXiv:hep-th/9309075].
  

\bm{kw} 
  W.~Kinnersley and M.~Walker,
  ``Uniformly Accelerating Charged Mass In General Relativity,''
  Phys.\ Rev.\ D {\bf 2} (1970) 1359.
  
\bm{exact}
See for example, 
 H.~Stephani, D.~Kramer, M.~MacCallum, C.~Hoenselaers and E.~Herlt,
  ``Exact solutions of Einstein's field equations,'' 
Cambridge, UK: Univ. Pr. (2003).


\bm{adsc0}
  J.~F.~Plebanski and M.~Demianski,
  ``Rotating, Charged, And Uniformly Accelerating Mass In General Relativity,''
  Annals Phys.\  {\bf 98} (1976) 98.

\bm{adsc1}
  O.~J.~C.~Dias and J.~P.~S.~Lemos,
  ``Pair of accelerated black holes in anti-de Sitter background: The AdS
  C-metric,''
  Phys.\ Rev.\ D {\bf 67} (2003) 064001
  [arXiv:hep-th/0210065].

\bm{dsc}
  O.~J.~C.~Dias and J.~P.~S.~Lemos,
  ``Pair of accelerated black holes in a de Sitter background: The dS
   C-metric,''
  Phys.\ Rev.\ D {\bf 67} (2003) 084018
  [arXiv:hep-th/0301046].
  
\bm{rs2} 
  L.~Randall and R.~Sundrum,
  ``An alternative to compactification,''
  Phys.\ Rev.\ Lett.\  {\bf 83} (1999) 4690
  [arXiv:hep-th/9906064].
  
\bm{pope1}
  H.~Lu and C.~N.~Pope,
  ``Branes on the brane,''
  Nucl.\ Phys.\ B {\bf 598} (2001) 492
  [arXiv:hep-th/0008050].

\bm{pope2}
  I.~Y.~Park, C.~N.~Pope and A.~Sadrzadeh,
  ``AdS braneworld Kaluza-Klein reduction,''
  Class.\ Quant.\ Grav.\  {\bf 19} (2002) 6237
  [arXiv:hep-th/0110238].


\bibitem{bousso}
  R.~Bousso,
  ``Positive vacuum energy and the N-bound,''
  JHEP {\bf 0011} (2000) 038
  [arXiv:hep-th/0010252].

\bibitem{tensor2}
  V.~Balasubramanian, J.~de Boer and D.~Minic,
  ``Mass, entropy and holography in asymptotically de Sitter spaces,''
  Phys.\ Rev.\ D {\bf 65} (2002) 123508
  [arXiv:hep-th/0110108].



\bm{cgh}
 A.~Chamblin, S.~W.~Hawking and H.~S.~Reall,
  ``Brane-world black holes,''
  Phys.\ Rev.\ D {\bf 61} (2000) 065007
  [arXiv:hep-th/9909205].

\bm{GL}
R.~Gregory and R.~Laflamme,
  ``Black strings and p-branes are unstable,''
  Phys.\ Rev.\ Lett.\  {\bf 70} (1993) 2837
  [arXiv:hep-th/9301052].

  
\bm{HM}
G.~T.~ Horowitz and K.~Maeda,
``Fate of the black string instability,''
Phys.\ Rev.\ Lett.\ {\bf 87} (2001) 131301 
[arXiv:hep-th/0105111].

\bm{hull0}  
  C.~M.~Hull,
  ``Timelike T-duality, de Sitter space, large N gauge theories and
  topological field theory,''
  JHEP {\bf 9807} (1998) 021
  [arXiv:hep-th/9806146].
  

\bibitem{brihaye}
  Y.~Brihaye and T.~Delsate,
  ``Black strings and solitons in five dimensional space-time with positive
  arXiv:hep-th/0611195.

\bibitem{BY}
  J.~D.~Brown and J.~W.~.~York,
   ``Quasilocal energy and conserved charges derived from the gravitational
  action,''
  Phys.\ Rev.\ D {\bf 47} (1993) 1407.

\bibitem{tensor1}
  V.~Balasubramanian and P.~Kraus,
  ``A stress tensor for anti-de Sitter gravity,''
  Commun.\ Math.\ Phys.\  {\bf 208} (1999) 413
  [arXiv:hep-th/9902121].


  

\bibitem{GW}
  G.~W.~Gibbons and S.~W.~Hawking,
  ``Cosmological Event Horizons, Thermodynamics, And Particle Creation,''
  Phys.\ Rev.\ D {\bf 15} (1977) 2738.

\bm{cosmo}
A.~M.~Ghezelbash and R.~B.~Mann,
  ``Entropy and mass bounds of Kerr-de Sitter spacetimes,''
  Phys.\ Rev.\ D {\bf 72} (2005) 064024
  [arXiv:hep-th/0412300].
\\
M.~H.~Dehghani and H.~KhajehAzad,
  ``Thermodynamics of Kerr Newman de Sitter black hole and dS/CFT
  correspondence,''
  Can.\ J.\ Phys.\  {\bf 81} (2003) 1363
  [arXiv:hep-th/0209203].



\bm{SW}
D.~Sudarsky and R.~M.~Wald,
  ``Extrema of mass, stationarity, and staticity, and solutions to the Einstein
  Yang-Mills equations,''
  Phys.\ Rev.\ D {\bf 46} (1992) 1453.
\\
 R.~M.~Wald,
  ``The First Law Of Black Hole Mechanics,''
  arXiv:gr-qc/9305022.


\bibitem{vio1}
  R.~Bousso, O.~DeWolfe and R.~C.~Myers,
  ``Unbounded entropy in spacetimes with positive cosmological constant,''
  Found.\ Phys.\  {\bf 33} (2003) 297
  [arXiv:hep-th/0205080].

\bibitem{vio2}
  R.~Clarkson, A.~M.~Ghezelbash and R.~B.~Mann,
   ``Entropic N-bound and maximal mass conjectures violation in four
  dimensional Taub-Bolt(NUT)-dS spacetimes,''
  Nucl.\ Phys.\ B {\bf 674} (2003) 329
  [arXiv:hep-th/0307059].

\bibitem{cai-ohta}
  R.~G.~Cai and N.~Ohta,
  ``Surface counterterms and boundary stress-energy tensors for  asymptotically
  non-anti-de Sitter spaces,''
  Phys.\ Rev.\ D {\bf 62} (2000) 024006
  [arXiv:hep-th/9912013].

\bm{cho-nam}
  J.~H.~Cho and S.~Nam,
  ``Living near de Sitter bubble walls,''
  arXiv:hep-th/0607098.


\bibitem{adsbr}
  H.~K.~Kunduri, J.~Lucietti and H.~S.~Reall,
  ``Do supersymmetric anti-de Sitter black rings exist?,''
  arXiv:hep-th/0611351.


\end{thebibliography}
\end{document}